\documentclass[prd,aps,floats,twocolumn]{revtex4}
\usepackage{epsfig}
\usepackage{times}
\usepackage{amsmath, amssymb}

\newcommand{\be}{\begin{equation}}
\newcommand{\ee}{\end{equation}}
\newcommand{\ba}{\begin{eqnarray}}
\newcommand{\ea}{\end{eqnarray}}
\newcommand{\bc}{}

\newcommand{\bra}[1]{\left(#1\right)}

\begin{document}

\preprint{\
\begin{tabular}{rr}
&
\end{tabular}
}
\title{Dark Matter, Modified Gravity and the Mass of the Neutrino.}
\author{P.~G~Ferreira$^{1}$,
C.~Skordis$^{2}$, C.~Zunckel$^{1}$}
%
\affiliation{
$^1$Astrophysics, University of Oxford, Keble Road, Oxford OX1 3RH, UK\\
$^{2}$ Perimeter Institute, Waterloo, Ontario N2L 2Y5, Canada
}

\begin{abstract}
It has been suggested that Einstein's theory of General Relativity
can be modified to accomodate mismatches between the gravitational
field and luminous matter on a wide range of scales. Covariant 
theories of modified gravity generically predict the existence of extra degrees of freedom which
may be interpreted as dark matter. We study a subclass of these theories where
the overall energy
density in these extra degrees of freedom is subdominant relative to
the baryon density and show that they favour the presence of massive
neutrinos. For some specific cases (such as a flat Universes with a cosmological constant) one finds a conservative lower bound on the neutrinos mass of $m_\nu>0.31$ eV.
\end{abstract}

\date{\today}
\pacs{PACS Numbers : }
\maketitle

\section{Introduction}
\noindent
There is compelling evidence that the baryons in the Universe
are unable to generate the gravitational potentials that we
observe on a wide range of scales.   A simple
paradigm can be used to explain this mismatch between light
and gravity: the Universe is filled with an appreciable amount
of matter which is cold (i.e. has non-relativistic velocities today)
and does not interact with light. It has been shown 
that Cold Dark Matter (CDM) can explain a host of observation,
from dynamics of clusters to the formation of the cosmic web \cite{DM1}.

The CDM paradigm has been proposed within the context of Newtonian gravity
and Einstein's theory
of General Relativity. It has been argued that these theories may not
be valid on all scales. Indeed, proposals for modifying gravity
 have been shown to fit much of the currently
available data \cite{MOND}.  A plethora of covariant theories
have been studied in detail; TeVeS gravity, 
modified Einstein-Aether theories,
conformal gravity, higher derivative actions, etc,  have been advocated as possible
rival theories to the CDM scenario \cite{Sanders,mannheim,Teves,Aether,Zlosnik,Zhao,gen_teves}.
There has been considerable effort in studying the 
cosmological consequences of these theories \cite{skordis,skordis_per,bourliot,Zlosnikcosmo}.
Given the level
of precision of current cosmological data, it is possible to
find severe constraints on these alternative theories and compare
their ability to describe nature with the CDM scenario.

There is an important, generic feature of covariant theories of modified 
gravity which is often overlooked: although they tamper with the gravitational
sector of the equations of motion, they also {\it inevitably} lead to the introduction of extra degrees of
freedom which may be interpreted as an exotic form of dark
matter. Let us exemplify. Theories which modify the Einstein-Hilbert
action by, for example, replacing the Ricci scalar, $R$, by a function of
different curvature invariants,
$f(R,R_{\alpha\beta}R^{\alpha\beta},\cdots)$, introduce higher derivative terms, and hence
new modes.
These new modes will contribute to
the overall energy density. This is patently obvious in the case of
theories where $f$ is simply a function of $R$; such theories can
be mapped onto normal Einstein gravity with an additional scalar field.
This also true of conformal gravity, where the action is now constructed
from the Weyl tensor. A field must be added to fix the scale of gravity
and the resulting low energy equations are fourth order \cite{FofR}.
More modern attempts at constructing theories of modified gravity have
the same characteristics in a much more explicit way. In TeVeS \cite{Teves}, a scalar
field and a vector field is introduced which not only modify the
gravitational field equations but also source the very same field
through their stress energy tensor. In generalized Einstein-Aether
theories, a time-like vector field is introduced \cite{Zlosnik}. 

Given what we have just said, there is an obvious question: aren't these
extra degrees of freedom simply a contrived form of dark matter?  
It is conceivable that the
extra degrees of freedom in modified theories of gravity may play 
such a role. If so, dark matter has been introduced through the back door.
It turns out that the role of extra degrees of freedom in theories of modified
gravity is more complicated than one might expect. In Skordis {\it et al} \cite{skordis},
it was shown that the extra degrees of freedom in TeVeS can make a negligible
contribution to the background (or overall) energy density. Indeed, if
TeVeS is to be consistent with big bang nucleosynthesis, the fractional
energy density in these extra degrees of freedom, $\Omega_X$, must be under
a percent.  Yet even though $\Omega_X\ll 1$,
fluctuations in the extra fields could have a significant impact on
the growth of structure. In particular, due to the modified nature of
 gravity, they could source the growth of gravitational potentials
and sustain them through Silk damping at recombination. These results
were corroborated in Dodelson and Liguori \cite{DL}, where the fluctuations in
the vector field were found to play an important role.

Hence some theories of modified gravity can fit current
observations of large scale structure, either from galaxy surveys or
the cosmic microwave background, even though $\Omega_X\ll1$. 
We would like to point
out that the latter property is not generic. In some incarnations
$\Omega_B\ll\Omega_X\simeq 1$ where $\Omega_B$ is the fractional energy
density in baryons. These theories end up being a hybrid of the two
paradigms, modified gravity and dark matter, and in principle
should be harder to distinguish from dark matter theories (although
there are some suggestions of specific tests) \cite{Zlosnik,Zhao}. 

In this paper we will try to expand on an
important feature of TeVeS pointed out in Skordis {\it et al} : if one assumes that
the Universe 
is flat and the only form of non relativistic matter consists of
$5\%$ baryons (consistent with Big Bang Nucleosynthesis), the angular
power spectrum of the Cosmic Microwave Background (CMB) will
differ significantly from observations. The only way to resolve this discrepancy is to introduce
some form of non-relativistic matter, and the only one allowed
within the known menagerie of fundamental constituents of the Universe 
is a massive neutrino. To match observations of the CMB, neutrinos
with a mass of approximately $2$ eV are needed. This 
result is clearly a hint of a more general statement that may
be made about theories of modified gravity in which the extra
degrees of freedom play a subdominant role: if these theoriess
are to agree with measurements of the CMB then they require the
presence of massive neutrinos. 
We wish to see if this implies a lower bound on the mass of
the neutrino.

\section{An approximate theory and cosmological observables}

Let us consider a generic modified theory of gravity in the limit of
homogeneity and isotropy. The {\it physical} metric (i.e. the metric
which is minimally coupled to the matter fields) can be parametrized
in terms of a scale factor, $a(t)$ which has a logarithmic derivative, $H=d\ln(a)/dt$. 
The energy density of the Universe can be 
split into the normal degrees of freedom, $\rho$ (such as baryons, photons,
neutrinos and dark energy) and the extra degrees of freedom that arise from the
modifications, $\rho_X$. The modified Friedman equations look somewhat
like
\begin{eqnarray}
F(a,H)H^2=\frac{8\pi G}{3}(\rho+\rho_X) \label{Friedmod}
\end{eqnarray}
where $G$ is Newton's constant and $FH^2$ is a function that arises from
varying the action for a particular theory. In fact it is convenient to rewrite the
equation in a more familiar form by defining an {\it effective} Newton's
constant $G_{eff}=G/F$. For the purpose of what follows we 
use a parametrization such that $G_{eff}\simeq G_{eff}(a)$; with
a sufficiently flexible choice of parameters we can encompass 
cases where $G_{eff}$ depends on $a$, $H$, etc.

We consider a sub class of the theories, in which $\rho_X<\rho_B$,
(where $\rho_B$ is the Baryon density). We consider  a parametrization
such that
\begin{eqnarray}
\rho_X\simeq f_{B}\rho_B+f_{R}\rho_R
\end{eqnarray}
where $\rho_R$ is the energy density in radiation. We have that $f_{B}<1$ and the correct abundance
of light elements requires that $f_{R}<10^{-2}$. We also include in $\rho$, a component
that behaves like dark energy, $\rho_{DE}$, with an equation of state $w<0$.
We find it convenient to parametrize the equation of state of the dark energy component as
$w=w_0+w_1z/(1+z)$.  Modifications to
the gravitational sector may lead to accelerated expansion at late times (such
as those proposed in \cite{mannheim,Zlosnik,bourliot,Zhao}), meaning that the dark energy could also arise from the extra fields in the modified gravity
sector. Our dark energy term includes all of these possibilities. 

With the evolution of the scale factor in hand, there are a few
observables that we may now calculate. Let us start off with
the position
of the first peak of the angular power spectrum of the Cosmic Microwave Background (CMB).
It is a
direct measure of the angular diameter distance and hence of
the expansion history of the Universe from recombination until
today and is the centrepiece of the analysis of this paper.
  Schematically we have the following picture \cite{HS}.
Before recombination (which occured at time $t_*$), photons
and baryons were tightly coupled and underwent acoustic oscillations.
During tight coupling the photon density contrast in the conformal Newtonian gauge,
 obeys the differential equation
\begin{equation}
 \ddot{\delta}_\gamma + \frac{3\rho_b}{3 \rho_b + 4 \rho_\gamma} \frac{\dot{a}}{a}\dot{\delta}_\gamma + k^2 c_s^2 \delta_\gamma = S[\Phi,\ddot{\Phi},\Psi]
\end{equation}
where $c_s^2 = \frac{4\rho_\gamma}{3(3\rho_b + 4\rho_\gamma)} $ is the sound speed, $S[\Phi,\ddot{\Phi},\Psi]$ is a source (a function of the gravitational potentials, $\Phi$ and $\Psi$) and derivatives are with conformal time $\tau$.
In the WKB approximation~\cite{HS} the two linearly independent solutions to the homogeneous part are $\cos{ k r_s}$ and $\sin{k r_s}$,
and depend on the sound horizon $r_s(\tau) = \int^\tau_0 c_s d\tau$. The important thing is that these solutions are valid for any theory of 
gravity for which photons and baryons see the same physical metric. All modifications are to gravity or additional fields
and  implicitely alter the inhomogenous part through the source term $S$
which only depends on the potentials $\Phi$ and $\Psi$ of that same physical metric.

The physical scale, $d_*$ of the acoustic waves is set by the
sound horizon at time $t_*$, i.e. $d_*\simeq \int_0^{t_*}c_s(t)dt$. 
After recombination,
photons decoupled from the baryons and freestreamed towards us,
travelling a distance given by $
d_0=ca_0\int_{t_*}^{t_0}dt/a=\int_{a_*}^{a_0}da(a^2H)$  where the subscript $0$ labels 
today. 
The angular
size on the sky of the sound horizon, $\theta_*$, is given by
$\theta_*\simeq \frac{a_0d_*}{a_*d_0}$
The sound horizon at last scattering leaves a very distinct
signature on the angular power spectrum of the CMB: a series
of peaks and troughs. The spectrum generated at recombination is related to the spectrum today via a projection through a spherical
Bessel function $j_\ell(k(\tau - \tau_*)$ , where $ad\tau=dt$, (an ultraspherical bessel function in the curved case).  Once again this is independent of the theory of gravity. The position of the peaks and troughs in the angular powerspectrum is primarily dependent on the
cosmological shift parameter, ${\cal R}$ which is related to the angular
diameter distance and is given by
\begin{eqnarray}
{\cal R}&=&\frac{1}{2}\sqrt{\frac{\Omega_{M}(a_0)}{\Omega_{K}(a_0)}}
\sin y \nonumber \\
y&=& \sqrt{|\Omega_K(a_0)|}\int_{a_*}^{a_0}\frac{da}{a^2(\sum_i\Omega_i(a))^{1/2}} \label{shift}
\end{eqnarray}
where $\Omega_i(a)$ is the fractional energy density of component
$i$ as a function of scale factor ($K$ corresponds to the curvature
and $M$ to non relativistc matter) \cite{efstathiou}.

Very few assumptions have gone into this calculation:
the Universe underwent recombination and the horizon structure is
a result of the expansion rate of the Universe.  
Current measurements of the CMB have reached a level of precision
such that it is practically impossible to deviate from this
simple picture \cite{WMAP5}. Attempts at changing these fundamental assumptions
inevitably lead to radical departures from this simple picture
and a gross mismatch to the data. So any theory of modified
gravity must lead to the basic picture of the CMB that we
infer from the data. Hence we can use our Eq. \ref{Friedmod} to work out ${\cal R}$ for theories
of modified gravity with the caveat that the fractional energy densities
must be rescaled by the effective Newton's constant, i.e. we must
replace $\Omega_i$ by $(G_{eff}/G_0)\Omega_i$ in Equation \ref{shift}.
Throughout this analysis we  consider a conservative bound on the
shift parameter: $ 1.63  <{\cal R}< 1.76 $ \cite{corasaniti}.

Another useful observable, as measured from the Hubble diagram of
distant supernovae, is the luminosity distance, $d_L$. It is related 
to the angular diameter distance, $d_A$, described above, through
$d_L=(1+z)d_A$, where $1+z=a_0/a$ defines the redshift at a given
value of the scale factor. While the CMB gives us one measure of
$d_A$ at $z\simeq1100$, the Hubble diagram of distant supernovae
gives us a series of measurements of $d_L$ out to $z \simeq 1.8$. Again,
as above we can use our modified Friedman equations to calculate
$d_L$ and compare to the data. Given that we do not have to use 
any information about perturbations about the background, we make
even fewer assumptions. We use the group of supernovae, termed the `gold' set, from the
HST/GOODS programme \cite{Riess:2004nr}, complemented by the
recently discovered higher redshift supernovae, reported in
\cite{Riess:2006fw}.

Finally, we  consider two more measurements. We  take into account the constraints on $\Omega_B$ from the
abundance of light elements. We  use $\Omega_Bh^2=0.022\pm 0.002$,
where the Hubble constant is defined to be $H_0=100h$km s$^{-1}$Mpc$^{-1}$.
Lastly we  consider current constraints from the Key Project
of the Hubble Space Telescope (HST) on the expansion rate today. We
 use $H_0=72\pm8$km s$^{-1}$Mpc$^{-1}$.

\section{Exploring Parameter space}

\begin{table}[t]
\begin{center}
\begin{tabular}{|c|c|c|}    \hline
\textbf{Model} & \textbf{68$\%$ CL} & \textbf{95$\%$ CL}
\\ 
\hline
$\Lambda$CDM  &  $  0.58 \leq m_{\nu} \leq 1.17$ & $ 0.31 \leq m_{\nu} \leq 1.48$ 
\\
$w$CDM &  $1.58 \leq m_{\nu} \leq 2.59$ & $0.92 \leq m_{\nu} \leq 3.02 $ 
\\
$w$CDM + $\Omega_{\kappa}$  &  $ 0.07 \leq m_{\nu} \leq 1.05$ & $m_{\nu} \leq 1.8$ 
\\
$\Lambda$CDM+$G(z)$  &  $0.018 \leq m_{\nu} \leq 0.62$ & $m_{\nu} \leq 1.04$ 
\\
$w$CDM + $G(z)$  &  $  1.05 \leq m_{\nu} \leq 2.33 $ & $ 0.22 \leq m_{\nu} \leq 2.68$ 
\\
$w$CDM +$\Omega_{\kappa}$ + $G(z)$ &  $m_{\nu} \leq 0.62$ & $m_{\nu} \leq 1.32 $ 
\\
$\Lambda$CDM +$G(\alpha, \gamma, z)$ & $ 0.034 \leq m_{\nu} \leq 0.71 $ & $m_{\nu} \leq 1.05$ 
\\
\hline
\end{tabular}
\end{center}
\caption{ Results for different cosmological models for a compilation data set. }
\label{results_SN}
\end{table}

We are interested in seeing if the presence of massive neutrinos is
a generic feature of the class of models that we are considering. 
We assume three families of neutrinos with identical masses,
and we  take the mass of each family, $m_\nu$, as the free parameter.
An obvious first case to study is a generalization of the TeVeS result
from Skordis {\it et al}, i.e. a Euclidean Universe with a cosmological constant
and a constant effective Newton's constant. Indeed we find that the
posterior for $m_\nu$ is positive, centered at $m_\nu\simeq 0.84$ eV and we
can set a lower bound on the neutrino mass at the 95$\%$ confidence
level (CL) of $m_\nu> 0.31$ eV, as shown in table (\ref{results_SN}). 
The value of $m_\nu$ proposed in Skordis {\it et al}
lies comfortably in that range $0.31<m_\nu<1.48$ eV.

Relaxing the assumption that the
acceleration is driven by a cosmological constant (i.e. freeing up
$w_0$ and $w_1$), leads to a lower bound of $m_\nu>0.92$ eV, slightly
stronger than in the previous case. 
The supernovae data strongly constrain the parameters
describing the nature of dark energy in the range $0 \leq z \leq 1.8$, the era where its contribution is dominant, 
and favour an effective $w(z) <-1$. This means that the contribution of the dark energy component
diminishes more rapidly than $\Lambda$ as a function of $z$.  To compensate, the neutrinos are 
required to be relativistic at the surface of last scattering and hence considerably more massive, 
giving rise to the observed shift in the distribution to higher masses.  The increased freedom 
gives a broader distribution. When the supernovae data set is removed, 
a larger contribution to the total energy from the dark energy component is allowed, weakening the lower bound on 
$m_\nu$, as shown in Table (\ref{results}).  However, for this simple class of theories, strong statements can now be made: 
there is a definite lower bound on the mass of the neutrino, as can be seen from Table (\ref{results_SN}).

Up until now, studies of theories of modified gravity have been undertaken
in the context of Euclidean Universes. Relaxing the assumption of spatial 
flatness greatly broadens the posterior distribution of $m_\nu$, in particular, extending it to as much
as $1.8$ eV at the 95$\%$ CL. These models correspond to closed Universes where $\Omega_K < 0$. 
In addition the dark energy parameters are less strongly peaked (due to the degeneracy with curvature).
Models in which the Universe is open are however favoured, leading to a weakened  lower bound 
of $m_\nu>0.07$ eV  but only at the $68\%$ CL. In the absence of the supernovae data, $\Omega_K$ is weakly 
constrained accordance with the increased freedom. This leads to a generally broader $m_\nu$ distribution with a similar peak.


We have been exploring the effect of the extra degrees of freedom
but we should expect modifications to the left hand side of equation \ref{Friedmod}.
We have parameterized this in terms of $G_{eff}$ that depends
on the scale factor. In principle, the time dependence of $G_{eff}$ can be more complex,
depending on the normal matter fields as well as the extra degrees of
freedom. Furthermore, for any given theory of modified gravity, the Bianchi
identities as well as the various couplings between $G_{eff}$ and the remaining sector,
impose specific constraints on its time evolution \cite{skordisnew}. I.e. we do no have
complete freedom to vary $G_{eff}$.

In what follows, we will be conservative and jettison any constraints that
come from consistency but we will consider two types of relatively general behaviour
which encompasse what we have found in a wide range of models.
One simple parametrization is
\begin{eqnarray}
G_{eff}=G_0(1+z)^n \nonumber
\end{eqnarray}
For example for TeVeS one finds that, for a sufficiently small $n$ one can adequately
mimic the behaviour of $G_{eff}$. Note that this parametrization does lead to
a monotonically changing $G_{eff}$, all the way back to recombination and so
must really only be considered an approximation- if not, it might lead to 
substantial changes to the peak structure in the CMB at recombination and we
have argued that this is clearly not the case.

\begin{figure}
\centering
  \includegraphics[trim = 0mm 0mm 0mm 0mm,scale = 0.7]{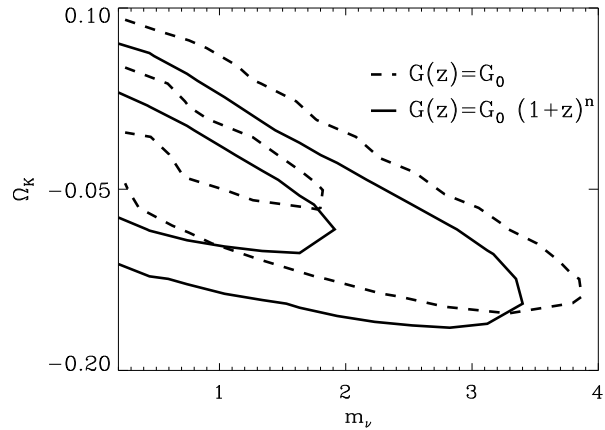} \\
  (a) \\
  \includegraphics[trim = 0mm 0mm 0mm 0mm,scale = 0.7]{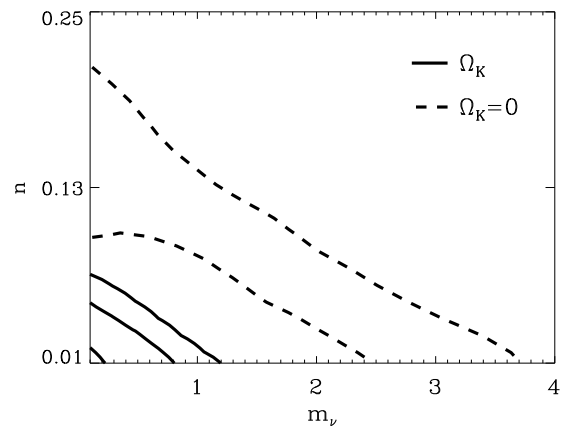}\\
  (b)
\caption{(a) The 68$\%$ and 95$\%$ confidence regions in $(m_\nu, \Omega_K)$ space from the compilation data set excluding the supernovae data. 
The region with dashed lines corresponds to a model where $n=0$ (ie. $G(z) = G_0$), while 
the region with solid lines corresponds to a Universe where $n \neq 0$ (ie. $G(z) =G(z, n)$).
(b) The 68$\%$ and 95$\%$ confidence regions in ($m_{\nu}, n$) space. The region with dashed lines corresponds to 
a model where $\Omega_K =0$, while the region with solid lines corresponds to a Universe where $\Omega_K \neq 0$. The plots illustrate the general trend of $G_{eff}$
to shift the distribution of $m_{\nu}$ to lower values, while 
$\Omega_K$ increases the range of neutrino masses that can be tolerated. } 
\label{contours}
\end{figure}

A variable $G_{eff}$ can have a substantial effect on allowed neutrino masses. In
particular, in a Euclidean Universe with cosmological constant, it 
lowers the required mass contribution from neutrinos significantly to $m_\nu< 0.35$ eV at the $2\sigma$ level such that the massless 
case is no longer ruled out.
This implies an anti-correlation between $n$ and $m_\nu$. 
Extending the model further to include dark energy again requires larger neutrino masses 
(lower bound of $0.22$ eV at 95$\%$ CL). 
However in non-flat Universe case, the freedom in parameter 
space means that a wide range of masses are tolerable, including the massless scenario and $m_{\nu}=1.32$ eV at the 2$\sigma$ 
level.  We note that allowing for the possibility of a time-dependent $G_{eff}(z)$ parameterized as above and admitting spatial curvature will
have similar effects on the Hubble equation. The primary effect of $G_{eff}$ is to shift the distribution of $m_{\nu}$ to lower values, while 
$\Omega_K$ increases the range of neutrino masses that can be tolerated. This is explicitly illustrated in Figure (\ref{contours}). The plot (a) 
compares the 68$\%$ and 95$\%$ confidence intervals in 
$(m_\nu, \Omega_K)$ space when $G_{eff}$ is time-independent (dashed lines) and dynamical (solid lines) and shows the shift in the allowed regions to more 
negative values of $\Omega_K$. 
Figure (\ref{contours}b) illustrates the impact of spatial curvature on constraints on $m_{\nu}$ 
in the presence of a time-dependent $G_{eff}$. The $1\sigma$ and $2\sigma$ 
regions are significantly reduced when curvature is admitted. 

Another possible parameterization is if $G_{eff}$ switches between two values
at some point in the past. For example, if $G_{eff}$ is approximately
six times larger during the baryon dominated era than it is now, the background
evolution will be essentially equivalent to that of dark matter dominated Universe
at that time. To mimic this effect we consider
\begin{eqnarray}
G_{eff}=G_0 \bra{1+\frac{\alpha z}{1+\gamma z}}
\end{eqnarray}
At low redshift, $G_{eff}$ starts at $G_0$ and increases linearly with $z$.  
At large redshift ($z >  1/\gamma$), G(z) tends towards a constant $~G_0 \alpha/\gamma$.
We limit the change in G(z) from $z=0$ to recombination by imposing the condition that $\alpha/\gamma < 5$ 
such that its does not change by factor of more than $6$.
We find that with this parametrization, which is reminiscent of a number of different models, that
the results are almost identical to that of the previous proposal for $G_{eff}$. Indeed, it is
the very late time behaviour of $G_{eff}$ that plays a significant role in changing the
observables and in that respects the two parametrizations are very similar.


\begin{table}
\begin{center}
\begin{tabular}{|c|c|c|}    \hline
\textbf{Model} & \textbf{68$\%$ CL} & \textbf{95$\%$ CL}
\\ 
\hline
$\Lambda$CDM  &  $ m_{\nu} \leq 0.16 $ & $  m_{\nu} \leq 0.38$ 
\\
$w$CDM &  $0.32 \leq m_{\nu} \leq 0.97$ & $0.064 \leq m_{\nu} \leq 1.27$
\\
$w$CDM + $\Omega_{\kappa}$  &  $ 0.02 \leq m_{\nu} \leq 1.27$ & $m_{\nu} \leq 3.065$
\\
$\Lambda$CDM +$G(z)$  &  $m_{\nu} \leq 0.14$ & $m_{\nu} \leq 0.35$ 
\\
$w$CDM + $G(z)$  &  $ 0.015 \leq m_{\nu} \leq 0.53$ & $ m_{\nu} \leq 0.92$
\\
$w$CDM +$\Omega_{\kappa}$ + $G(z)$ &  $m_{\nu} \leq 1.38$ & $m_{\nu} \leq 2.79$
\\
$\Lambda$CDM +$G(\alpha, \gamma, z)$ & $m_{\nu} \leq 0.16$  & $m_{\nu} \leq 0.35$
\\
\hline
\end{tabular}
\end{center}
\caption{ Results for different cosmological models for a compilation 
data set where the supernovae data is excluded. }
\label{results}
\end{table}

\section{Discussion}
It has been claimed that modified theories of gravity inevitably require the presence 
of massive neutrinos and that these may be sufficiently massive to be measurable
with up and coming neutrino experiments  such as KATRIN \cite{KATRIN}. This claim has
been triggered by two pieces of anecdotal evidence. Firstly that the simplest TeVeS
model needs neutrinos to fit the angular power spectrum of the CMB as shown
in \cite{skordis}. And secondly, that attempts at reconciling observed and inferred masses
of clusters requires the presence of a massive neutrino halo \cite{sanders, famaey}.
In this letter we have attempted to extend the remit of the first piece of evidence.

We have found that, although for a restricted set of models, we can place a lower bound on the mass of
the neutrino, for more general ranges of parameters, it is possible to satisfy
the subset of cosmological constraints without having to invoke massive neutrinos.
This is not to say that specific models with, for example, a variable effective Newton's
constant might not lead to a tight constraint on the neutrino mass. But it is
clearly not possible to make a definitive statement on the mass of the neutrino
for general theories of modified gravity. Theories must be studied case by case and
we have shown how this can be done in an economical way.

It may be possible to come up with constraints on the neutrino masses from a different
set of observables, related to the second piece of evidence. For example, 
in the simplest picture of a cluster in these
theories, neutrinos seem to be inevitable to be able to make up dynamical
mass measurements and weak lensing observations. This simple picture is
incomplete and much of the work that has been done on clusters in the
context of modified gravity has opted to ignore the extra degrees of freedom \cite{Dai}.
They can play a significant role and, in the same way as for large scale observations,
may substantially weaken cluster constraints on the neutrino mass. A more
detailed analysis of these systems must be undertaken before definitive
conclusions can be inferred.

{\it Acknowledgments}:
We thank A. Cooray, A. Melchiorri, G. Starkman and T. Zlosnik for discussions.
C. Zunckel is supported by a Domus A scholarship awarded by Merton College. 
Research at Perimeter Institute for Theoretical Physics is supported in part by the Goverment of Canada through
NSERC and by the Province of Ontario through MRI.


\end{document}